%% file: proceedings.tex
\pdfoutput=1
\documentclass[a4paper, twoside, openany]{book}

\usepackage[svgnames]{xcolor} 
\usepackage{setspace}
\usepackage{pdfpages}
\usepackage[hidelinks]{hyperref}
\usepackage{xspace}
\usepackage[bottom=1in, top=1in, left=32mm, right=38mm]{geometry}

\usepackage{fancyhdr}
\usepackage{tocloft}

\makeatletter
\let\latexl@section\l@section
\def\l@section#1#2{\begingroup\let\numberline\@gobble\latexl@section{#1}{#2}\endgroup}

\let\latexl@chapter\l@chapter
\def\l@chapter#1#2{\begingroup\let\numberline\@gobble\latexl@chapter{#1}{#2}\endgroup}
\makeatother

\newcommand{\paper}[4]{%
\includepdf[pages=-, pagecommand={},addtotoc={1, chapter, 1, {#4\\\protect\emph{#3}}, paper:#1}]{#2}%
}

\newcommand{\crule}[3][black]{\textcolor{#1}{\rule{#2}{#3}}}
\definecolor[named]{lipicsYellow}{rgb}{0.99,0.78,0.07}
\newcommand{\chbox}[0]{\crule[lipicsYellow]{0.65cm}{0.65cm}}

\begin{document}

\frontmatter

\pagestyle{empty}

\input{title}

\cleardoublepage

\pagestyle{plain}
\pagenumbering{roman}
\setcounter{page}{1}


\input{preface}

\input{organization}


\newpage
\tableofcontents


\mainmatter
\pagestyle{fancy}
\pagenumbering{arabic}
\setcounter{page}{1}

\addtocontents{toc}{\vspace{0.5cm}}
\addtocontents{toc}{\textbf{\large Invited Talk}\par}

\chapter*{\chbox\xspace Invited Talk}
\newpage

\paper
{0}
{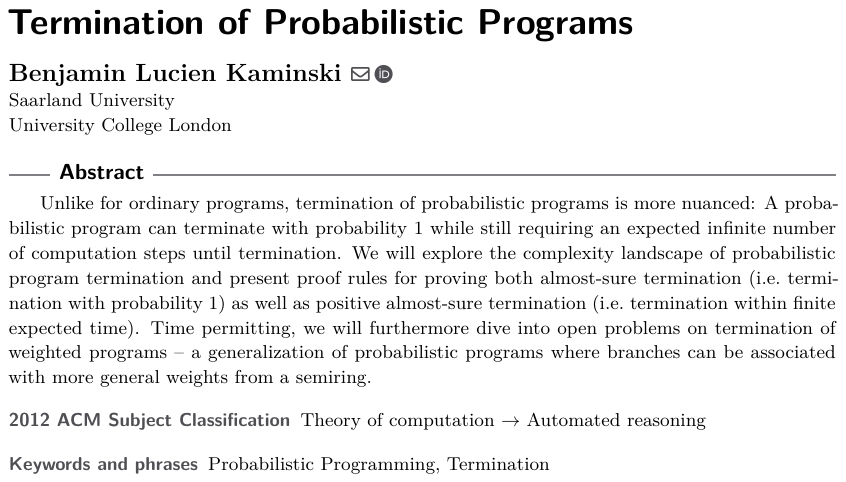}
{Benjamin Lucien Kaminski}
{Termination of Probabilistic Programs}

\addtocontents{toc}{\vspace{0.5cm}}
\addtocontents{toc}{\textbf{\large Regular Papers}\par}

\newcommand\arXiv[3]{%
\addtocontents{toc}{\vspace{0.5cm}}
\addtocontents{toc}{{#2\\\emph{\mbox{}\hspace{1em} #3}}\protect\cftdotfill{\cftdotsep}\href{https://doi.org/10.48550/arXiv.#1}{arXiv:#1}\par}
}

\arXiv{2307.14094}
{On Singleton Self-Loop Removal for Termination of LCTRSs with Bit-Vector Arithmetic}
{Ayuka Matsumi, Naoki Nishida, Misaki Kojima, Donghoon Shin}

\arXiv{2307.13973}
{Generalizing Weighted Path Orders}
{Teppei Saito, Nao Hirokawa}

\arXiv{2307.10061}
{Automated Complexity Analysis of Integer Programs via Triangular Weakly Non-Linear Loops (Short WST Version)}
{Nils Lommen, Eleanore Meyer, Jürgen Giesl}

\arXiv{2307.09839}
{Proving Non-Termination by Acceleration Driven Clause Learning (Short WST Version)}
{Florian Frohn, Jürgen Giesl}

\arXiv{2307.11549}
{Binary Non-Termination in Term Rewriting and Logic Programming}
{Étienne Payet}

\arXiv{2307.14671}
{A Verified Efficient Implementation of the Weighted Path Order}
{René Thiemann, Elias Wenninger}

\arXiv{2307.10002}
{Dependency Tuples for Almost-Sure Innermost Termination of Probabilistic Term Rewriting (Short WST Version)}
{Jan-Christoph Kassing, Jürgen Giesl}

\arXiv{2307.11024}
{Automated Termination Proofs for C Programs with Lists (Short WST Version)}
{Jera Hensel, Jürgen Giesl}

\arXiv{2307.13519}
{Higher-Order LCTRSs and Their Termination}
{Liye Guo, Cynthia Kop}

\arXiv{2307.14036}
{Hydra Battles and AC Termination, Revisited}
{Nao Hirokawa, Aart Middeldorp}

\arXiv{2307.14149}
{Old and New Benchmarks for Relative Termination of String Rewrite Systems}
{Dieter Hofbauer, Johannes Waldmann}

\arXiv{2307.14805}
{Linear Termination over N is Undecidable}
{Fabian Mitterwallner, Aart Middeldorp, René Thiemann}

\arXiv{2307.13426}
{Complexity Analysis for Call-by-Value Higher-Order Rewriting}
{Cynthia Kop, Deivid Vale}

\chapter*{\chbox\xspace Tool Descriptions}
\newpage

\addtocontents{toc}{\vspace{0.5cm}}
\addtocontents{toc}{\textbf{\large Tool Descriptions}\par}

\paper
{1}
{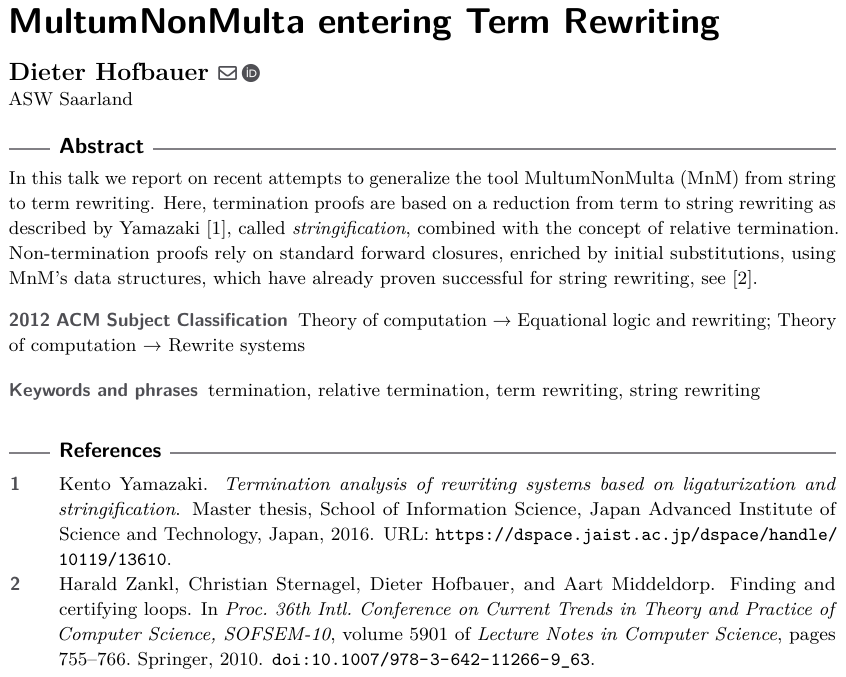}
{Dieter Hofbauer}
{MultumNonMulta entering Term Rewriting}

\paper
{2}
{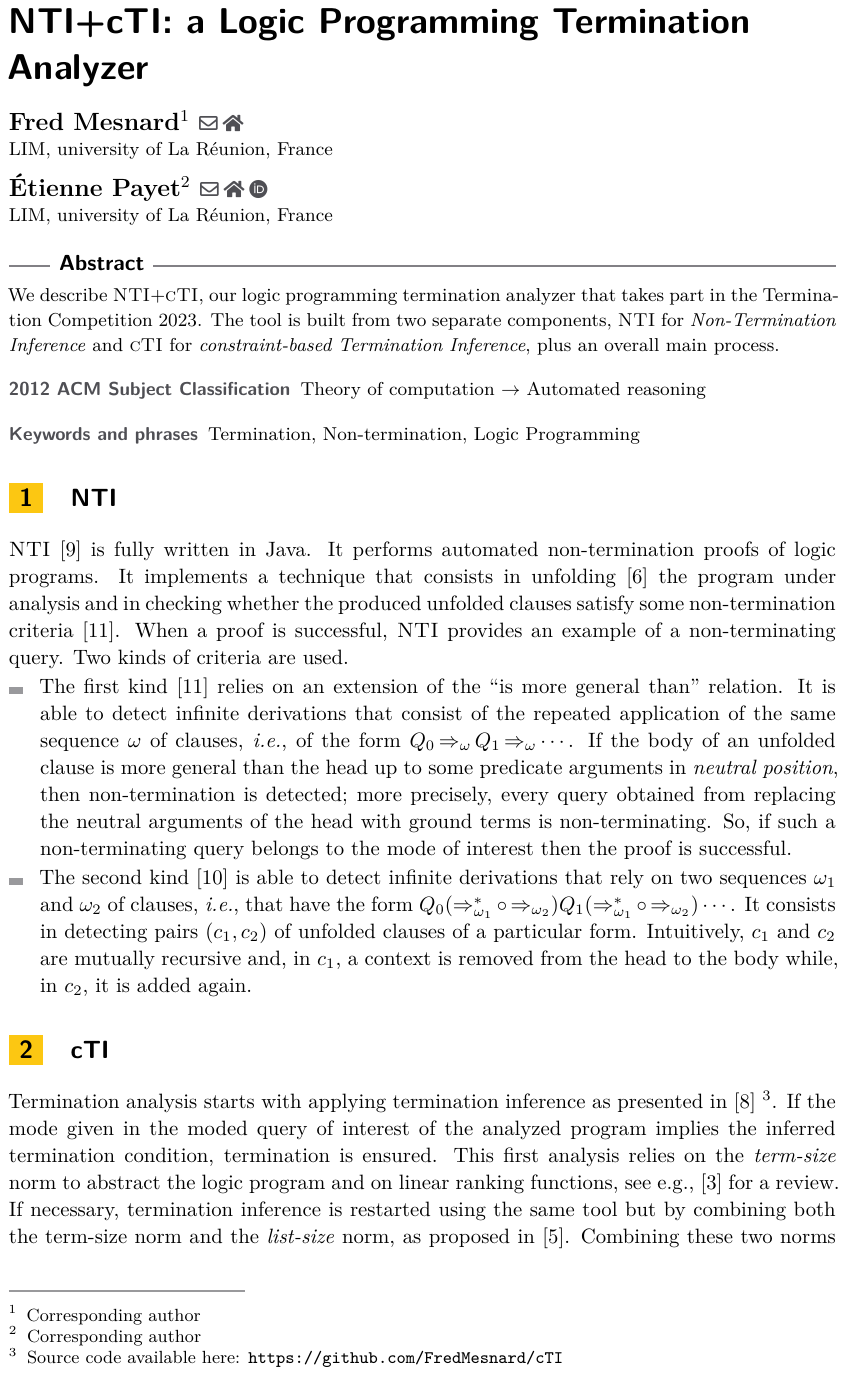}
{Fred Mesnard and \'Etienne Payet}
{NTI+cTI: a Logic Programming Termination
Analyzer}

\end{document}

%% file: title.tex

\begin{titlepage} 

\begin{spacing}{2}

  \vspace*{3cm}
  {\huge
    
  \noindent
  \textsf{\textbf{19th International Workshop on}}

  \noindent
  \textsf{\textbf{Termination}}
  }
\end{spacing}

  \vspace*{1.5cm}
  \noindent
  {\large
    \textsf{\textbf{WST 2023, August 24--25, 2023}}}
  \par\noindent
  {\large\textsf{\textbf{held as part of Obergurgl Summer on Rewriting 2023}}}
  \par
  \vspace*{0.5cm}
  \noindent
  {\large
    \url{https://termination-portal.org/wiki/WST2023}
  }

\vspace*{3.0cm}

\begin{spacing}{2}
  \noindent
  \textsf{\large Edited by}~\\
  \textsf{\Large Akihisa Yamada}

\vspace*{9cm}
\noindent
\hfill\colorbox{lipicsYellow}{\Large \textsf{\textbf{\textcolor{black!70}{WST 2023~~~~~~~~~~~~~~~~~~~~~~~~~~Proceedings~~~~~~~~~~~~~~~~~~~~~~~~~~~~~~~~~~~~~~~~~}}}}

\end{spacing}
  \end{titlepage}

%% file: preface.tex

\chapter*{\chbox\xspace Preface}
\addcontentsline{toc}{chapter}{Preface}

This report contains the proceedings of the 19th International
Workshop on Termination (WST 2023), which was held in Obergurgl during
August 24--25 as part of Obergurgl Summer on Rewriting (OSR 2023).

\medskip
The Workshop on Termination traditionally brings together, in an
informal setting, researchers interested in all aspects of
termination, whether this interest be practical or theoretical,
primary or derived. The workshop also provides a ground for
cross-fertilization of ideas from the different communities interested
in termination (e.g., working on computational mechanisms, programming
languages, software engineering, constraint solving, etc.). The
friendly atmosphere enables fruitful exchanges leading to joint
research and subsequent publications.
The 19th International Workshop on Termination continues the
successful workshops held in St. Andrews (1993), La Bresse (1995), Ede
(1997), Dagstuhl (1999), Utrecht (2001), Valencia (2003), Aachen
(2004), Seattle (2006), Paris (2007), Leipzig (2009), Edinburgh
(2010), Obergurgl (2012), Bertinoro (2013), Vienna (2014), Obergurgl
(2016), Oxford (2018), the virtual space (2021), and Haifa (2022).

\medskip
The WST 2023 program included an invited talk by
Benjamin Lucien Kaminski
on
\emph{Termination of Probabilistic Programs}.
WST 2023 received 13 regular submissions and six abstracts for tool
presentations, two of which were accompanied by a system description.
After light reviewing the program committee decided to accept all
submissions.
The proceedings is published on arXiv. Each of the 13 regular
submissions is made available as an arXiv article, and the two system
descriptions are combined into this preface, due to the arXiv policy
not to accept very short papers.

\medskip

I would like to thank the program committee members
for their dedication and effort, and the organizers of
OSR~2023 for the invaluable help in the organization.

\vspace*{1cm}

\noindent
Innsbruck, August 2023 \hfill Akihisa Yamada

%% file: organization.tex

\chapter*{\chbox\xspace Organization}
\addcontentsline{toc}{chapter}{Organization}

\section*{Program Committee}

\bigskip
\begin{tabular}{l@{\qquad}l}
Martin Avanzini& INRIA Sophia Antipolis\\
Florian Frohn& RWTH Aachen\\
Carsten Fuhs& Birkbeck, U. London\\
Raúl Gutiérrez& U. Politécnica de Madrid\\
Étienne Payet& U. La Réunion\\
Albert Rubio& Complutense U. Madrid\\
René Thiemann& U. Innsbruck\\
Deivid Vale& Radboud U. Nijmegen\\
Johannes Waldmann& HTWK Leipzig\\
Akihisa Yamada (chair)& AIST Tokyo Waterfront
\end{tabular}

%

\bigskip